\documentclass[12pt,iop]{emulateapj}

\usepackage{natbib}
\usepackage{epsfig,graphicx,times}
\usepackage{subfigure}

\DeclareGraphicsExtensions{.eps,.eps.gz,.epsi,.ps}

\shorttitle{Thermal emission for a WASP-43b in $H$ and $K_{\rm s}$-bands}
\begin{document}
\title{Ground-based detections of thermal emission from the dense hot Jupiter WASP-43b in
 $H$ and $K_{\rm s}$-bands
}

\author{W Wang$^{1}$
, R van Boekel$^{2}$, N Madhusudhan$^{3}$,  G Chen$^{4}$, G Zhao$^{1}$, Th Henning$^{2}$}

\address{
$^{1}$Key Laboratory of Optical Astronomy, National Astronomical Observatories,
Chinese Academy of Sciences, Beijing 100012, China
}
\address{
$^{2}$Max Planck Institute for Astronomy, K\" onigstuhl 17, D-69117 Heidelberg, Germany}
\address{
$^{3}$Department of Physics and Department of Astronomy, Yale University, New Haven, CT 06511, USA. }

\address{
$^{4}$ Purple Mountain Observatory, Chinese Academy of Sciences, West Beijing Road, Nanjing 210008, China
}

\email{wangw@nao.cas.cn}

\begin{abstract}
We report new detections of thermal emission from the transiting hot Jupiter WASP-43b in the $H$ and $K_{\rm s}$-bands as observed at secondary eclipses. The observations were made with the WIRCam instrument on the CFHT. We obtained a secondary eclipse depth of 0.103$_{-0.017}^{+0.017}\%$ and 0.194$_{-0.029}^{+0.029}\%$ in the $H$ and $K_{\rm s}$-bands, respectively. The $K_{\rm s}$ band depth is consistent with 
previous measurement in the narrow band centered at 2.09$\mu$m by \cite{Gillon2012}.  Our eclipse depths in both bands are consistent with a blackbody spectrum with a temperature of $\sim1850$\,K, slightly higher than the dayside equilibrium temperature without day-night energy redistribution. Based on theoretical models of the dayside atmosphere of WASP-43b, our data constrain the day-night energy redistribution in the planet to be $\lesssim 15-25$\%, depending on the metal content in the atmosphere. Combined with energy balance arguments our data suggest that a strong temperature inversion is unlikely in the dayside atmosphere of WASP-43b. However, a weak inversion cannot be strictly ruled out at the current time. Future observations are required to place detailed constraints on the chemical composition of the atmosphere.

\end{abstract}
\keywords{stars: planetary systems - stars: individual: WASP-43 - techniques: photometric}
\maketitle

\section{Introduction}

Observations of thermal emission from transiting exoplanets during secondary eclipse provide important constraints on the thermal structure and chemical composition of their dayside atmospheres. Such observations have been reported for over 30 exoplanets to date~\citep{SD2010}, including hot Jupiters 
(e.g. \citealt{Charbonneau1995}), hot Neptunes \citep[e.g.][]{Stevenson2010}), and even super-Earths \citep{Demory2012}, using a wide range of facilities from space, such as {\it Spitzer}~\citep[e.g.][]{Knutson2008, Fressin2010} and {\it HST}~\citep[e.g.][]{Swain2009a, Swain2009b}, and from ground~\citep[e.g.][]{Sing2009, Croll2010}.  Concomitant theoretical efforts towards interpreting such data have led to the inferences of molecular species \citep{Swain2009a, MS2009}, thermal inversions
 \citep{Burrows2008, Machalek2008, Knutson2009a, MS2010}, thermal phase curves \citep{Knutson2009b}, non-equilibrium chemistry \citep{Stevenson2010, MS2011, Moses2011}, carbon-rich atmospheres \citep{Madhusudhan2011b,
Madhusudhan2011a, Madhusudhan2012b}, and several classification schemes for irradiated giant planets \citep{Fortney2008, Knutson2010, Cowan2011, Madhusudhan2012c}. 

In the recent past, it has become feasible to observe thermal emission from hot Jupiters using ground-based infrared telescopes of various sizes \citep[e.g.][]{Sing2009a,Croll2011}. Most hot Jupiters have their spectral energy distribution (SED) peak in the near-infrared \citep{Barman2008}, making them particularly good candidates to observe using ground-based near-infrared instruments. Such observations from ground have been reported in several infrared photometric channels  ranging from $z^\prime$ \citep[0.9 $\mu$m; e.g.][]{Smith2011} to K-band \citep[2.1\,$\mu$m;][]{Deming2012}. By combining such near-infrared ground-based data with space-borne data obtained with {\it HST}~\citep[$\sim 1 - 2\, \mu$m;][]{Swain2012} and {\it Spitzer}~\citep[between 3.6 - 24\,$\mu$m; ][]{Knutson2012}, a long spectral baseline is obtained, allowing one to place stringent constraints on the thermal profiles and chemical compositions of the atmospheres \citep[e.g.][]{Madhusudhan2011a}. Furthermore, ground-based photometric channels that are devoid of strong molecular absorption serve as excellent opacity windows in the exoplanetary SEDs, thereby probing the thermal profiles in their deep atmospheres \citep{Madhusudhan2012b}.

In the present work, we observe thermal emission in the $H$-band (1.6 $\mu$m) and $K_s$-band (2.1 $\mu$m) from the hot Jupiter WASP-43b from ground. WASP-43b \citep[][hereafter H11]{Hellier2011} is a hot Jupiter transiting a very cool K7-type dwarf \citep[$T_{\rm eff}=4520\pm120$,][hereafter G12]{Gillon2012}. The system has a special place in the parameter space of both stellar and planetary characteristics. The host star is known to be one of the lowest mass stars ($M=0.72\pm0.03\,M_{\odot}$) hosting a transiting planet, thereby leading to a large radial-velocity amplitude (H11). The star is young ($<0.1$\,Gyr), as derived from its lithium abundance, and is active as indicated by the presence of Ca\,HK lines. The rotation period of the star is $\sim 15.6\pm0.4$\,d with a significance of $>99.9$\%.  

The planet has a period of 0.81\,d and orbits at a separation of 0.01526\,AU, making it the closest-orbiting hot jupiter known. Secondly, the planet has a mass of $2.03\pm0.05\,M_{\rm Jup}$ and a radius of $1.04\pm0.02R_{\rm Jup}$, and therefore a density of $1.83\rho_{\rm Jup}$, i.e., 2.42\,${\rm g\,cm}^{-3}$ (G12). We note that WASP-43b is $\sim50\%$ more compact than Neptune, the solar system giant with the highest density. Among the 248 transiting planets with radii and masses determined, the mean density is $\sim0.72\rho_{\rm Jup}$. Among all the hot Jupiters whose atmospheres have been observed, WASP-43b has the second highest density, after the very massive planet XO-3\,b\,\citep{Johns-Krull2008}. The unusually high density of WASP-43b favors an old age and a massive core (G12) in the planet. Furthermore, the close orbital separation places WASP-43b amongst the highly irradiated class of hot Jupiters which are expected to host thermal inversions according to theory \cite[][but cf. \citealt{Spiegel2009}]{Fortney2008}. However, the high stellar activity may also mean that any inversion causing compounds may have been photochemically destroyed, thereby precluding the formation of a thermal inversion (Knutson et al. 2010). In yet another possibility, if the atmospheric composition of WASP-43b is carbon-rich inversion-causing compounds such as TiO and VO may be naturally absent \citep{Madhusudhan2011a, Madhusudhan2012b}. Thus, given the various peculiarities of the system, the planet WASP-43b forms an interesting candidate for detailed atmospheric characterization.

In what follows, we first describe our observations and data reduction process in Section~2. In Section~3, we present our detailed data analysis and results. We interpret our data and compare them to model spectra in Section~4. We summarize our work in the last section. 

\begin{figure}
\label{fig:density}
\includegraphics[width=8.5cm,clip=true,bb=5 10 400 460]{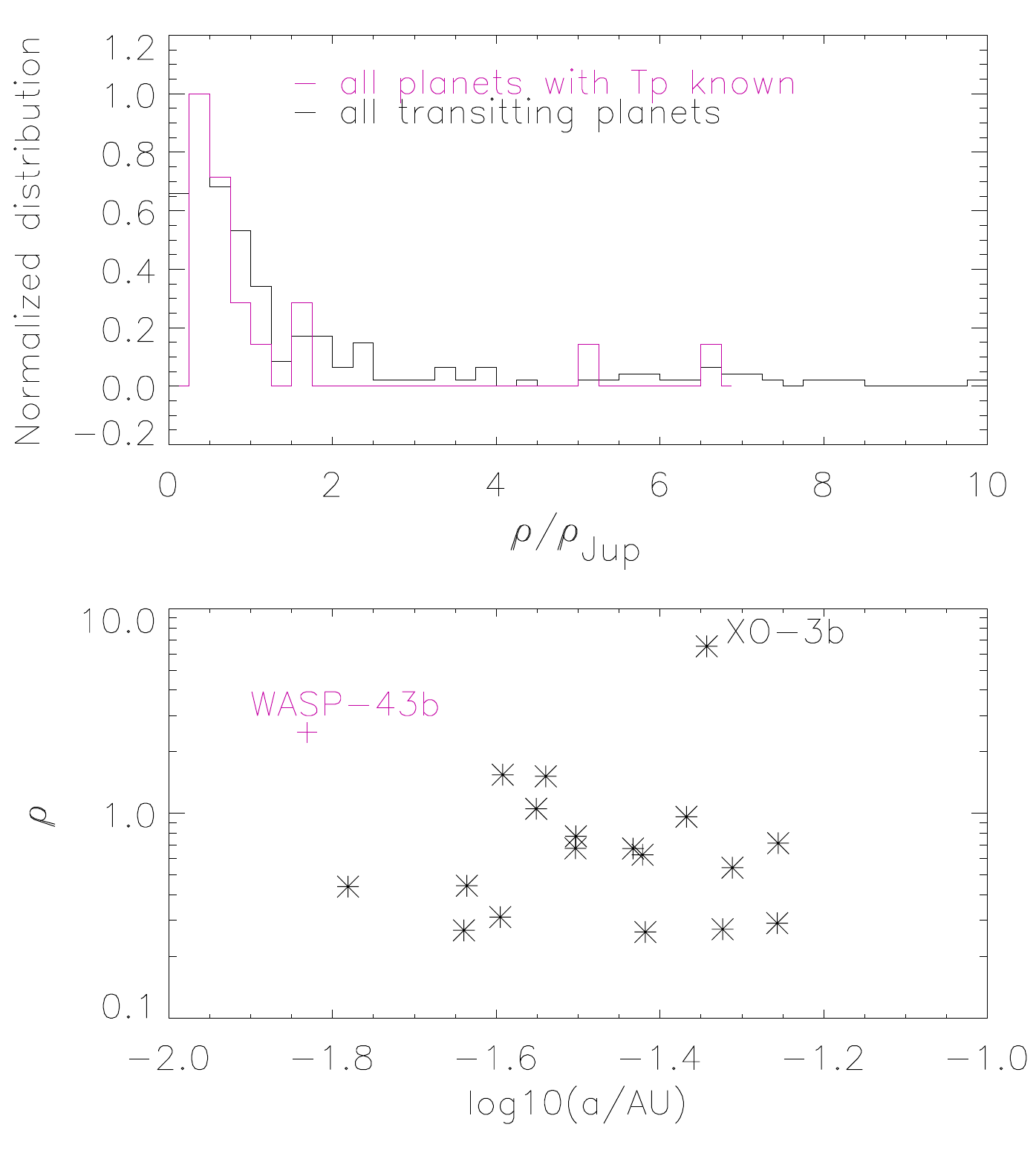}
 \caption{{\it Top} Normalized distribution of densities of all transiting planets (black line) and
 of all planets with thermal temperature known from secondary eclipse measurements in the 
 literature (red line); 
 {\it Bottom}\, planet density vs.  semi-major axis for all the planets that with atmospheres studies. 
 Data taken from the extrasolar planets Encyclopaedia (http://exoplanet.eu).
}
\end{figure}


\section{observations}

WASP-43b was observed twice during a secondary eclipse with WIRCam on the CFHT. We observed it in the $K_{\rm s}$ band on 2012 March 2 and in the $H$ band on 2012 April 10. For each night, we covered the entire eclipse and 1 hour of ``baseline'' before and after the eclipse. The airmass remained below 1.4 and 2.2, respectively, for the two observing nights. Conditions on March 2 were photometric, with a median seeing of 0.54\arcsec and a median total extinction of 0.08\,mag. Conditions on April 8 were less good, with a median seeing of 1.04\arcsec and a median total extinction of 0.14\,mag.

Both observations were carried out in Staring mode \citep{Devost2010}, which has proven to yield the best photometric precision~\citep{Croll2010}. We use the full WIRCam array with its 21$'\times21'$ field of view (FOV). The science observations lasted for $\sim$3.3 hours for each night, with the observing window centered in time on the predicted mid-eclipse time, assuming a circular orbit. A sky model was made by observing a dither pattern with 15 offsets prior and after the photometric time series, and taking the median of these to remove the stars. The sky model is scaled to the background level of each individual exposure of the photometric time series and subtracted.

We defocused the telescope to 1.9 and 2.0\,mm for the $H$- and $K_{\rm s}$-band observations, respectively, to minimize the impact of flat field errors and intra-pixel sensitivity variations, as well as to keep the flux of the target star well below the detector saturation level. The flux of our target and of the reference stars was therefore spread over a ring of $\sim$19 pixels (6\arcsec) in diameter on the CCD. We used an integration time of 5 seconds (DIT\,$=$\,5\,s) and averaged 12 integrations (NDIT\,$=$\,12) before writing the result to disk in order to increase the overall efficiency. In this way, we achieved an overall duty cycle of 42\%.

\begin{figure*}
\centering
\label{fig:fov}
\includegraphics[width=12cm,clip=true,bb=1 1 427 427]{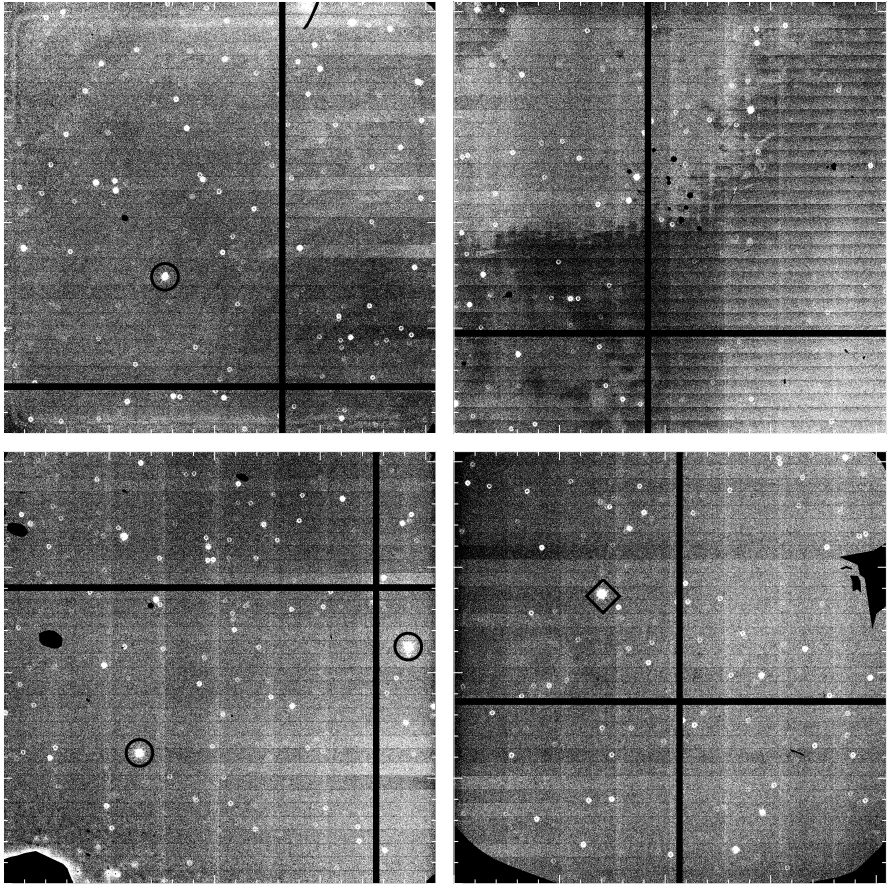}
 \caption{CFHT/WIRCam full frame reduced image during our observation of WASP-43b in the $K_{\rm s}$-band. The black vertical and horizontal bars in each chip are the columns and rows through which the guide stars (located under the intersection) were read out at high frequency. The target star, WASP-43 is marked with a diamond and the three reference stars that were finally used are marked with circles.}
\end{figure*}

In Fig.~\ref{fig:fov} we show the field that was observed. The science target and photometric reference stars that were used for $K_{\rm s}$ are marked with a diamond and circles, respectively. Because WASP-43 is very bright the number of useful reference stars is limited. This leads to relatively large uncertainties, compared to theoretical predictions from \citet{Devost2010}. The raw data show obvious artifacts from bad pixels, and the columns and rows that were used for the fast readout of $5\arcsec\times5\arcsec$ window for the guiding star in each chip.

\section{data analysis and results}

\begin{figure}
\label{fig:flux}
\includegraphics[width=8.5cm,clip=true,bb=31 54 545 770]{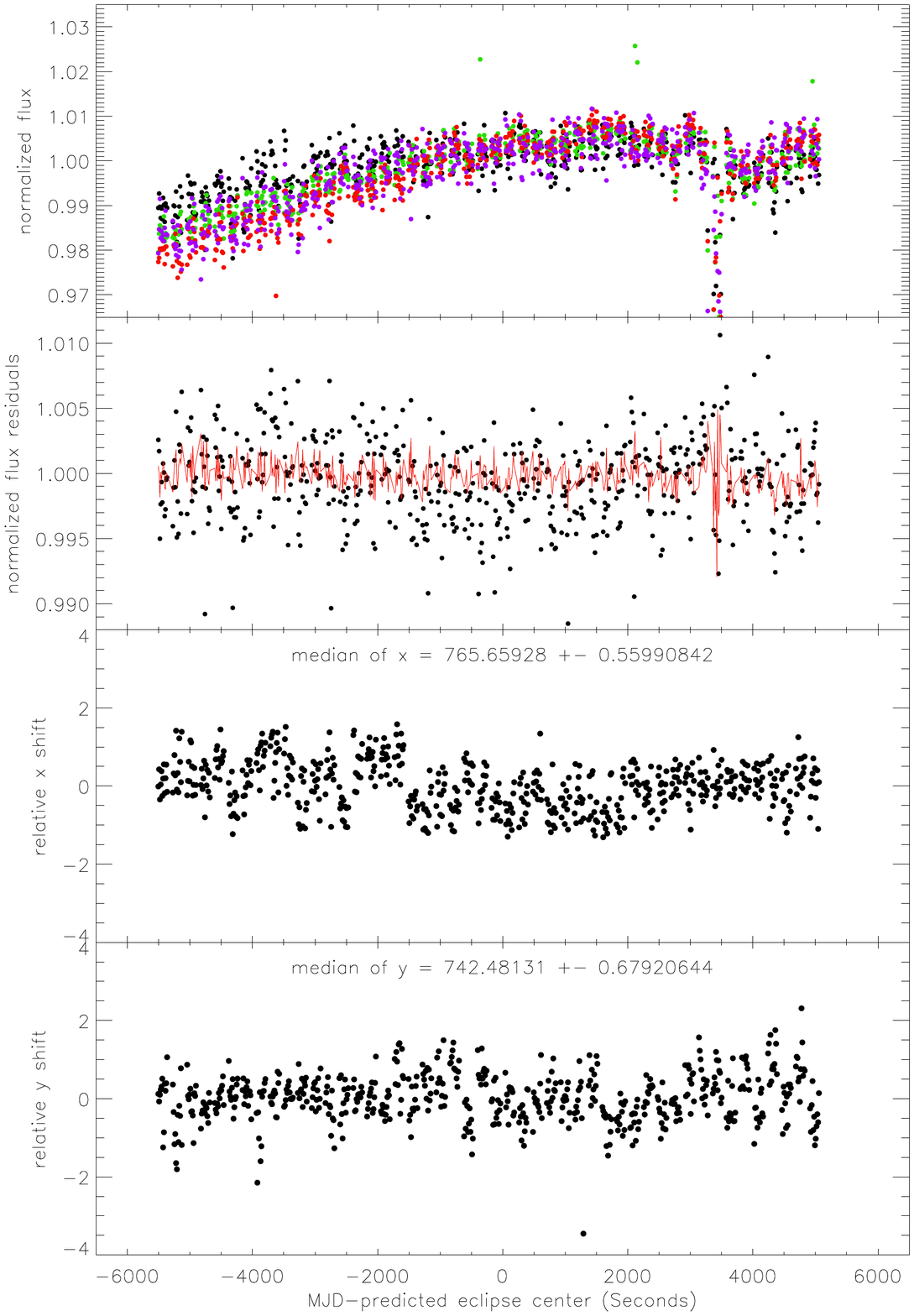}
 \caption{The reduced fluxes and relative positions of WASP-43's centroid in $K_{\rm s}$ as an example.  Top panel presents the normalized fluxes of all 4 stars, including the target and the 3 reference stars.
 The second panel shows the final WASP-43 fluxes (black dots) normalized with the weighted median of 
 all reference fluxes (the red line). The bottom two panels show the relative position of WASP-43's centroid.
}
\end{figure}

The raw data were reduced using $'$I$'$iwi pipeline{\footnote{http://www.cfht.hawaii.edu/Instruments/Imaging/WIRCam/IiwiVersion1Doc.html}}, including the following steps: nonlinearity flux correction, masking bad and saturated pixels, dark subtraction, flat fielding, and sky subtraction. 
Aperture photometry was performed on the reduced images using IDL programs for the target star and dozens of ``candidate'' reference stars which are bright but not saturated. We applied 8 different aperture sizes with a step of 0.5\,pixels, and determined that an aperture with a diameter of 17.5 pixels gives 
the smallest out-of-eclipse scatter in the H-band light curve. The corresponding ``optimal'' aperture diameter is 16 pixels for $K_{\rm s}$. Varying the apertures by $\pm$1 pixels in diameter results in consistent eclipse depths with, however, larger error bars.

\subsection{$K_{\rm s}$-band Observations}

We use the technique presented by \citet{Everett2001} to obtain the final high-precision differential aperture photometry. We first assemble a group of reference stars whose out-of-eclipse light curve variations are similar to that of the science target. We then create an average reference light curve by taking the flux-weighted mean, which is then used to normalize the light curve of the science target.

The top panel of Fig.\ref{fig:flux} shows the extracted fluxes of the target and three reference stars, normalized to their median values. The sudden jump at $\sim$3500\,seconds is probably caused by a rise of extinction by 0.05\,mag. WASP-43 is very bright ($K{\rm s}=9.27$\,mag) and there are only 8 stars of similar brightness in the field of view. ($\Delta m<2.5$\,mag). Three of these yield out-of-eclipse light curves whose shapes resemble that of the science target. In the rest 5 stars, two are used as guide stars or are too close to guiding windows and therefore had to be rejected. The remaining 3 stars are the faintest reference candidates, and were further excluded due to the large statistical uncertainties on their fluxes. The following panel presents the flux of WASP-43b after subtracting a median of the three reference stars. We emphasize that using 1, 2 or 3 of the three reference stars yields very similar differential light curve with mean
absolute differences $<6.0\times10^{-4}$, however with larger variations of ``baseline'' fluxes around the 3500\,s jump for the cases of 
fewer reference stars. The pointing stability and the weather conditions were good and we did not find a well-defined correlation between photometry and centroid positions of WASP-43.

\begin{figure}
\includegraphics[width=8.5cm,clip=true,bb=5 5 550 245]{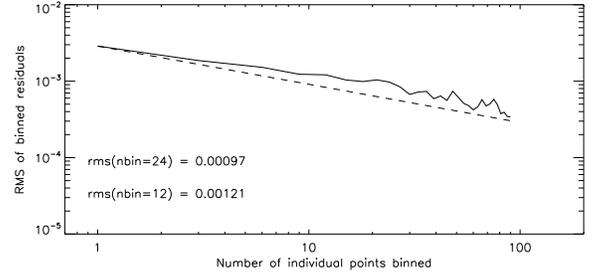}
\caption{\label{fig:rms} Comparison of WASP-43 noise level with Gaussian expectation after background correction as a function of binned points. The sold line present the noise levels of the actual data, while the dashed line indicates the Gaussian noise expectation. }
\end{figure}

An obvious background trend is observed in the normalized light curve of WASP-43 (Top panel of Fig\,\ref{fig:lc}). We applied a linear regression of the out-of-eclipse light curve with time in the form: 
\begin{equation}
B_{f}=c_{1}+c_{2}\,t,
\end{equation}
where $c_{1}$ and $c_{2}$ are fit parameters, $t$ is the time relative to the eclipse center in seconds.
This correction improves the point-to-point scatter of our out-of-eclipse data from a rms of $2.5\times10^{-3}$ to $1.2\times10^{-3}$ per every 60\,s  integration (or 12 images). Note that this is still far above the photon noise limit for a 60\,s integration on our target star, which is $\sim1.7\times10^{-4}$ in $K_{\rm s}$. Given that most of known data sets of exoplanets display these background trends, it is unlikely that this slope is intrinsic to WASP-43. 
We examine the noise level in our WASP-43 photometry,  after background correction, as a function of the temporal bin size (i.e. number of images averaged) in Fig.\,\ref{fig:rms}. The residuals are approximately consistent with the $\Delta t^{-1/2}$ behavior expected for normally distributed noise.

We first modelled our secondary transit light curves using the publicly available IDL routine suite EXOFAST$\footnote{http://astroutils.astronomy.ohio-state.edu/exofast}$\citep{Eastman2012}. This package uses Markov Chain Monte Carlo (MCMC) methods \citep{Ford2005} to fit the \citet{Mandel2002} model transit light curves to observations .

In our case, we enforced the limb darkening coefficients to be zero to realize the ``secondary eclipse'' case.  The EXOFAST light curve model has a total of 12 parameters.
We ran $10^{6}$ steps in the MCMC chain, yielding  best-fit results that are shown in the first column of Table~\ref{tbl:tapmcmc}. 

To examine the statistical significance and robustness of the fitting, and to estimate the errors on the derived eclipse depths, a standard bootstrap method~\citep{Press1992} was applied to the light curve. In each bootstrapping iteration, we uniformly resample the data with a random replacement, which yields a duplicate rate of $1/e$, i.e.,  $\sim$36.7\% for each sample. 1000 iterations are performed, and we obtain a Gaussian distribution of the best-fit eclipse depths, with a median and 1$\sigma$ standard deviation of $0.179\pm0.026\%$ for the $K_{\rm s}$ band, well consistent with the best fit value from the original light curve.

Using the EXOFAST model, designed for primary eclipse, to fit a secondary eclipse light curve has some side effects. In a primary eclipse, the eclipse depth is given by the square of the planet/star size ratio, with a small correction for orbital inclination if the star has appreciable limb darkening at the wavelength of observation. The duration of ingress and egress depends linearly on the planet size, and is hence coupled to the eclipse depth. In secondary eclipse, the eclipse depth depends on the square of the planet/star size ratio, multiplied by the ratio of the average surface brightness of star and planet. This yields much smaller eclipse depths, which EXOFAST mimics by a very small planet radius. This causes the inferred duration of ingress and egress to be much smaller than the actual duration, and hence the functional shape of the light curve model is not correct during ingress and egress, which makes the fitting improperly.

Therefore, we developed our own IDL package EXOPIKA for the light curve fitting, employing PIKAIA\citep{Charbonneau1995}, a general purpose optimization routine based on a genetic algorithm. We use the same input parameters to produce a model light curve, and to find a best fit to the observed curve. This code is faster than EXOFAST by a factor of $\sim$10 to achieve similar $\chi^{2}$ for the case of secondary eclipse, but does not yield error bars.
For a robustness check, bootstrap iterations were run 1000 times, yielding mean fitting results well consistent with those of original light curve.  The standard deviations of every parameters are taken as error bars. As expected, small differences between the results from EXOFAST and EXOPIKA exist, which are well within the uncertainties, as shown in Table~\ref{tbl:tapmcmc}.

\begin{table*}
\caption{Best-fit secondary eclipse parameters in $K_{\rm s}$ and $H$.}
\centering
\label{tbl:tapmcmc}
\begin{tabular}{lcccccccc}
\hline
\hline
\noalign{\smallskip}
Parameter & \multicolumn{4}{c}{$K_{\rm s}$} & \multicolumn{4}{c}{$H$}\\
\noalign{\smallskip}
\hline
\noalign{\smallskip}
          & EXOFAST & bootstrap & EXOPIKA & bootstrap & EXOFAST & bootstrap & EXOPIKA & bootstrap\\
\noalign{\smallskip}
\hline
\noalign{\smallskip}

$\Delta F(\%)   $&                 0.180$^{+0.031}_{-0.030}$    & 0.179$^{+0.026}_{-0.026}$    & 0.194   &0.196$_{-0.029}^{+0.029}$    & 0.102$^{+0.024}_{-0.023}$    & $0.101_{-0.020}^{+0.020}$     &0.103   &$0.102_{-0.017}^{+0.017}$\\
       $a/R_{*}$ &                 4.969$^{+0.082  }_{-0.076} $ & 4.971$_{-0.002}^{+0.002}$    & 5.142   &5.107$_{-0.233}^{+0.233}$    & 4.969$^{+0.083}_{-0.075}$    & $4.967_{-0.006}^{+0.006}$     &4.525   &$4.639_{-0.27}^{+0.27}$\\
\noalign{\smallskip}
t$_{\rm off}$(min)&                0.994$^{+1.181  }_{-1.253} $ & 0.720$_{-0.547}^{+0.0547}$   & 1.917   &1.440$_{-2.718}^{+2.718}$    & 1.244$^{+1.440}_{-1.339}$    & $0.769_{-0.720}^{+0.720}$     &7.073   &$7.148_{-2.521}^{+2.521}$ \\ 
\noalign{\smallskip}
$T_{\rm c}$\tablenotemark{a}(day)&88.3833$^{+0.0008}_{-0.0009}$ & 88.3831$^{+0.0037}_{-0.0037}$& 88.3839 &88.3836$_{-0.0019}^{+0.0019}$&127.4315$_{-0.0010}^{+0.0009}$& $127.4312_{-0.0005}^{+0.0005}$&127.4355&$127.4356_{-0.0018}^{+0.0018}$ \\
\noalign{\smallskip}
  Reduced $\chi^{2}$ &      3.26                                &                              & 3.17    &                             & 1.93                         &                               & 1.88\\
\noalign{\smallskip}
\noalign{\smallskip}
\hline
\noalign{\smallskip}
\end{tabular}
\begin{list}{}{}
\item[$^{a}$] The secondary eclipse center $T_{\rm c}$, in units of HJD-2,455,900. Offset due to light travel time is corrected.
\end{list}
\end{table*}

\subsection{$H$-band Observations}

The situation in $H$ is somewhat more complicated. Due to bad weather conditions on 2012 April 10, we only found one 
proper reference star (2MASS\,10200126-0948099, $H$=9.71), which has a light curve similar to that of WASP-43. The brighter reference star BD-09\,3048 ($H=9.23$) saturated in dozens of frames and therefore could not be used. In addition, close to time of mid-eclipse, the pointing jumped suddenly by $\sim2$\,\arcsec \ in DEC and $\sim$0.5\,\arcsec \ in RA, which has an obvious impact on the final light curve and was corrected in post-processing.

The $H$-band data underwent the same data reduction and analysis procedures as the  $K_{\rm S}$-band data. In addition, we performed a de-correlation of the light curve by fitting the out-of-eclipse photometric flux to the relative $x$ and $y$ positions of the centroid of the PSF with the function: 
\begin{equation}
f=1+a_{1}x+a_{2}y+a_{3}xy,
\end{equation}

where $a_{1}$, $a_{2}$ and $a_{3}$ are constants. This step is also performed in other similar studies such as \citet{Croll2011} and \citet{Zhao2012}. This correction is applied to the whole set of photometry, which improves the point-to-point scatter of our out-of-eclipse data from an RMS of $1.6\times10^{-3}$ to $1.1\times10^{-3}$ per every 60\,s (or 12 images). A background correction, identical to that done in the $K_{\rm s}$ band, further reduces the scatter to $0.74\times10^{-3}$. The light curves are fitted in an identical fashion to the $K_{\rm s}$ band data, yielding best fit parameters listed in Table~\ref{tbl:tapmcmc}.

\subsection{Other parameters}

While performing the light curve fits, we fix the period to the value derived by G12. The derived offsets from the expected eclipse center are: $1.9$ and $7.1$\,minutes for $K_{\rm s}$ and $H$ respectively. This is consistent with a small eccentricity of $\sim$0.0035, as given in G12. Our best fit inclination angle and eccentricity both match the values given in G12.

\subsection{Eclipse Depths and Brightness Temperatures}

\begin{figure}
\centering

\includegraphics[width=8.5cm,clip=true,bb=5 5 550 810]{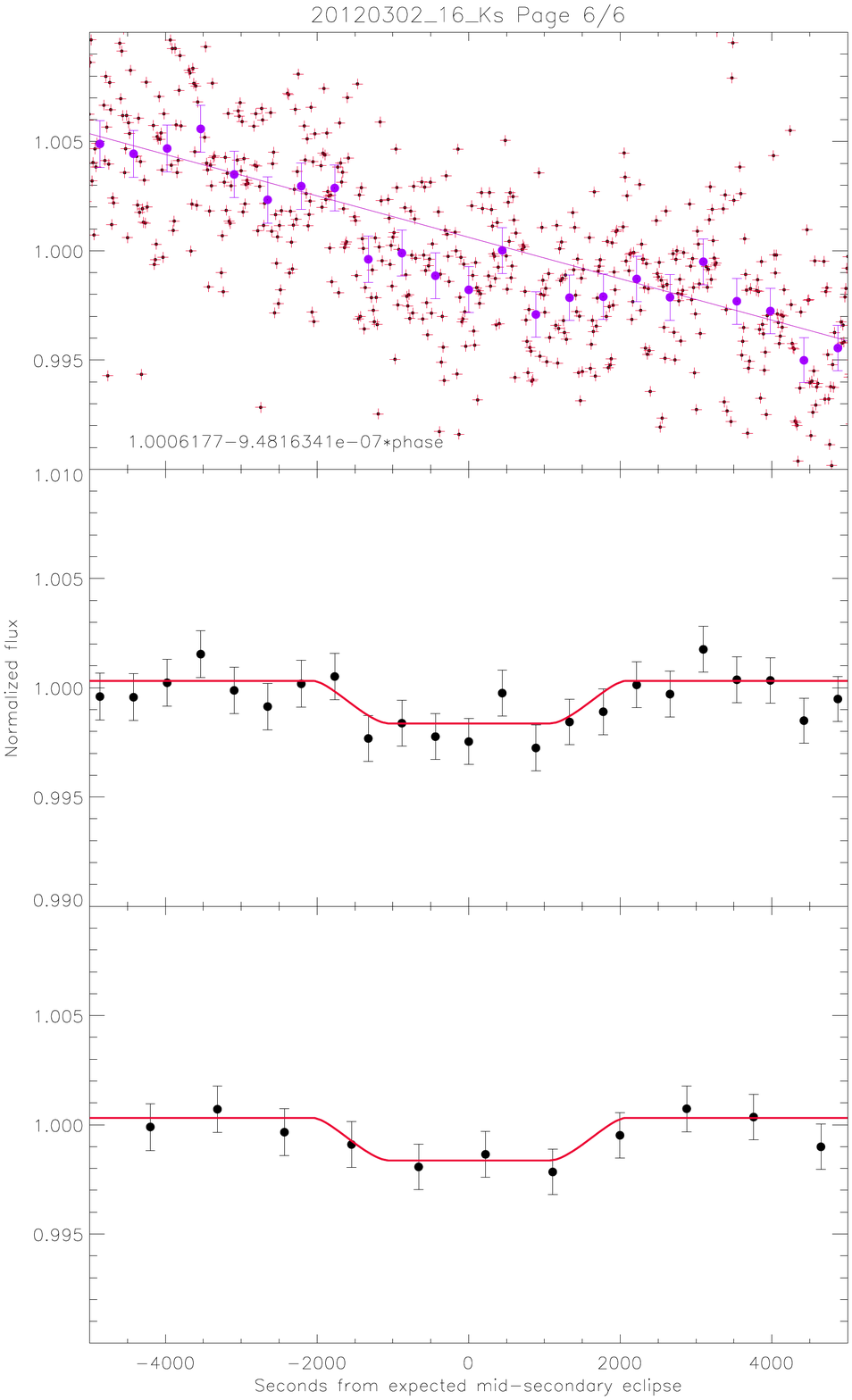}
 \caption{Light curve of WASP-43b in $K_{\rm s}$ as an example. Top panel shows the curve before background subtraction, with solid
 line for the background model. Middle and bottoms panels present the final corrected light curves with
 best-fit model in solid lines. The filled circles with error bars are rebinned data with bin sizes of 24 and 48, respectively. }
 \label{fig:lc}
\end{figure}

The depths of our best-fit secondary eclipses are 0.194\% and 0.103\% (cf. Fig~\ref{fig:lc}) with reduced $\chi^{2}$ of 3.2 and 1.9 respectively. The $\chi^{2}$ values suggests that residual systematic effects are present that are not removed in the data processing and are not included in light curve model. Such effects may include chromatic changes in the Earth atmospheric extinction that affect stars differently depending on their intrinsic spectral shape. Also, the the guide stars have changed several times during the $K_{\rm s}$ observation, leading to variations of spatial positions of stars on the detector, which leads to small apparent photometric variations if the flatfield correction is imperfect.

The depth we determined for $K_{\rm s}$ suggests a brightness temperature of $T_{B}=1718_{-107}^{+108}$\,K, assuming the planet and the star radiate as blackbodies and adopting a stellar effective temperature of $4520\pm120$\,K, a stellar radius of $0.667_{-0.011}^{+0.010}$\,R$_{\odot}$, and a planet radius of $1.036\pm0.019$\,R$_{\rm Jup}$ (G12).
This is higher than the equilibrium temperature of WASP-43b of $T_{\rm eq}=1434_{-46}^{+45}$\,K, assuming isotropic re-radiation and a zero Bond albedo. The eclipse depth in $H$ corresponds to a brightness temperature of $T_{\rm B}=1760_{-132}^{_+128}$\,K, consistent with that of $K_{\rm s}$ band.


\citet{Gillon2012} obtained eclipse depths of $0.156\,^{+0.015}_{-0.016}$\%  and $0.082\,^{+0.035}_{-0.037}\%$
in the narrow-band filters NB2090 centered  at 2.09$\mu$m and NB1190 at 1.19$\mu$m, respectively. 
The former value agrees  with our $K_{\rm s}$  measurement within uncertainties. Their theoretical models, when combined with thermal emission in these two narrow-band filters, indicate no day- to night-side redistribution of absorbed energy, no appreciable opacity of TiO and VO, and no thermal inversion. Their best fit model matches our $H$-band eclipse depth.

\section{The atmosphere of WASP-43b}

\begin{figure*}
\plottwo{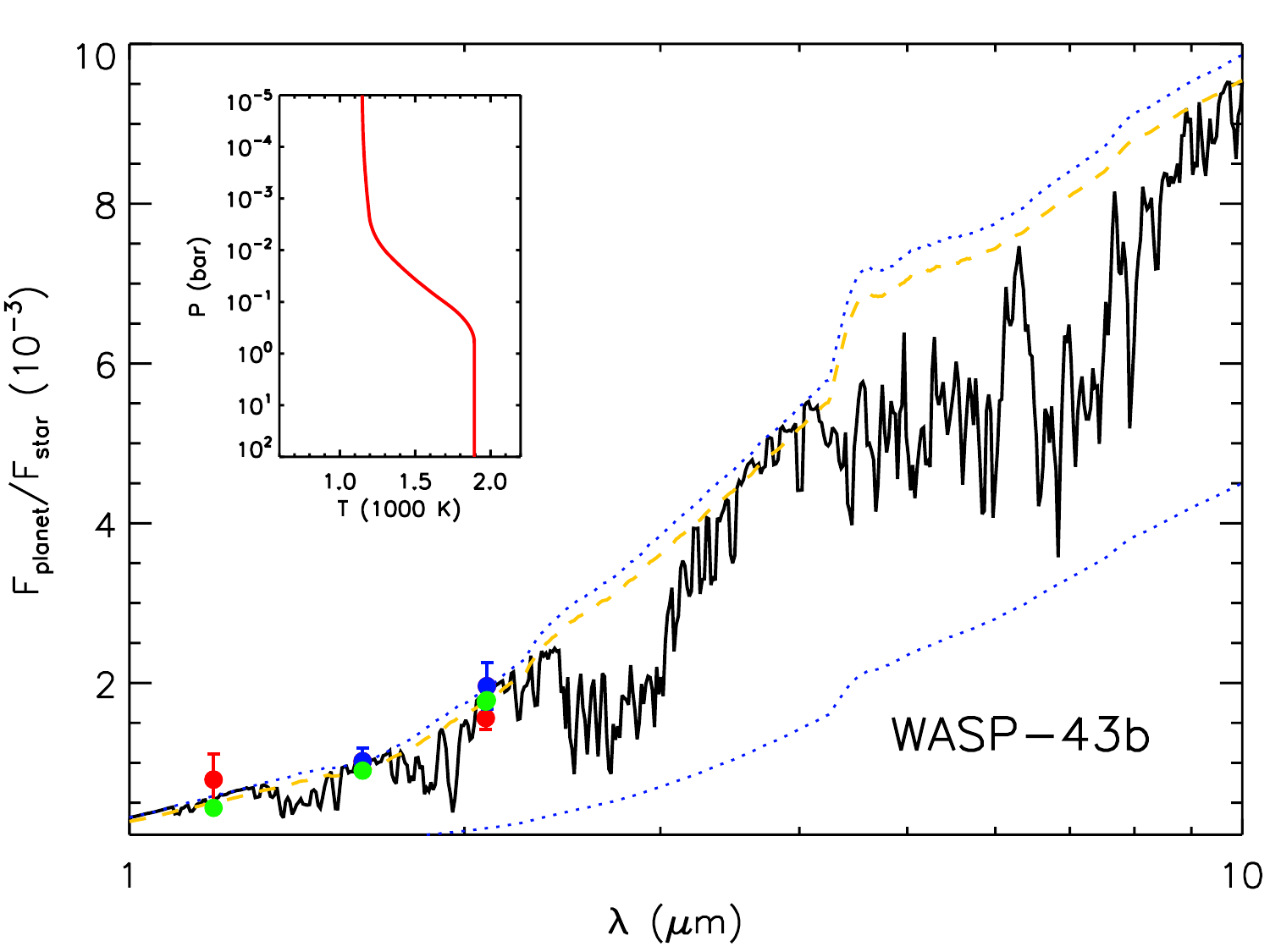}{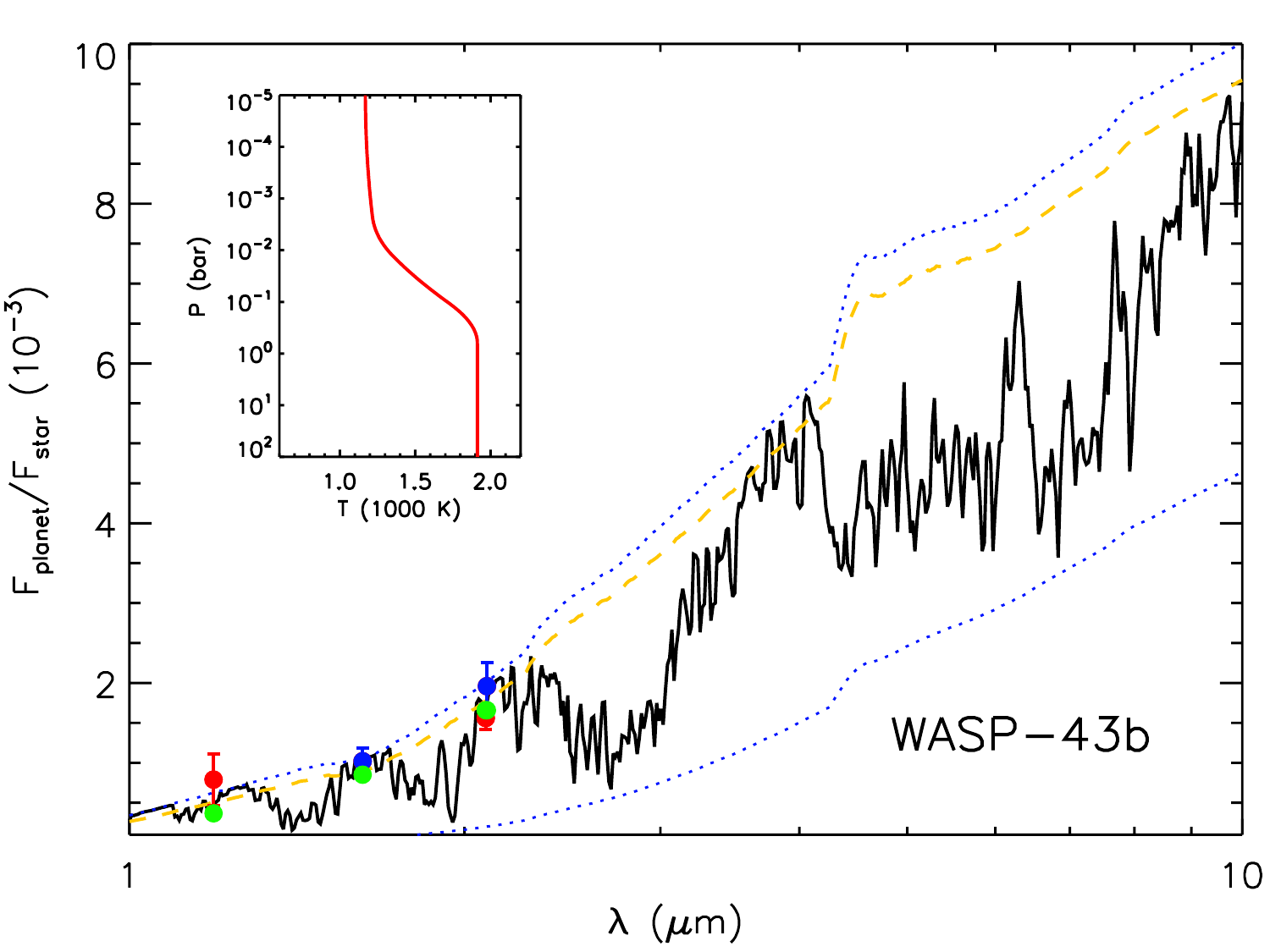}
 \caption{Observations and model spectra of dayside thermal emission from WASP-43b. In each of the two panels, the filled circles with error bars show the observed data. The blue data points show our broadband photometric observations in the $H$ and $K_s$ bands centered at 1.6 $\mu$m and 2.1 $\mu$m, respectively. The red data show previously published narrow band photometric data at 1.19 $\mu$m and 2.09 $\mu$m from \citet{Gillon2012}. The black curve shows a model spectrum and the red curve in the inset shows the corresponding temperature profile. The green filled circles show the model binned to the same resolution as the data. The blue dotted curves show blackbody spectra corresponding to the lowest and highest temperatures in the temperature profile. The orange dashed curve shows a blackbody spectrum corresponding to 1850 K. The two panels show two different model spectra with similar temperature profiles but different chemical compositions : (a) Left: model atmosphere with solar metallicity, and (b) Right: model atmosphere with 5$\times$ solar metallicity.} 
\label{fig:sed}
\end{figure*}

We place constraints on the atmospheric properties of WASP-43b by comparing our data with model spectra. We model the dayside thermal emission from WASP-43b using the exoplanetary atmospheric modeling and retrieval technique of \citet{MS2009}. The model computes line-by-line radiative transfer in a plane-parallel atmosphere in local thermodynamic equilibrium (LTE), and assumes hydrostatic equilibrium and global energy balance. The molecular mixing ratios, i.e. the chemical composition, and the pressure-temperature ($P$-$T$) profile of the 1-D atmosphere are input parameters to the model. The model atmosphere includes the major sources of opacity expected in hot hydrogen-dominated atmospheres, namely, molecular absorption due to H$_{2}$O, CO, CH$_{4}$, and CO$_{2}$, and continuum opacity due to H$_{2}$-H$_{2}$ collision-induced absorption (CIA). Our molecular line-lists are discussed in \citet{MS2009, Madhusudhan2012b}. Given observations of thermal emission from the planet, we explore the space of atmospheric chemical composition and temperature structure to determine the regions in model space that best fit, or are excluded by, the data \citep[see e.g.][]{Madhusudhan2011a}. In the present case, however, the number of available data points ($N = 2$) are far below the number of model parameters ($N = 10$), implying that a unique model fit to the data is not feasible. Consequently, we nominally fixed the chemical compositions of the models to fiducial abundances, such as solar elemental abundances in thermochemical equilibrium \citep[e.g.][]{BS1999, Madhusudhan2012b}, and explored the space of thermal profiles, with and without thermal inversions, that explain the data.  

Our observations constrain the thermal structure of the lower atmosphere of WASP-43b. Since the photometric $H$ and $K_s$ bands are relatively devoid of strong molecular absorption, compared to the {\it Spitzer} IRAC bands \citep[see e.g.][]{Madhusudhan2011a}, observations in these bands probe the deepest layers in the atmosphere from where emergent flux travels relatively unimpeded to the observer. For highly irradiated giant planets, the temperature profile in the optically thick regime of the dayside atmosphere is expected to be isothermal \citep{Hansen2008, Guillot2010}. As such, the brightness temperatures observed in the $H$ and $K_s$ bands constrain the isothermal temperature structure at the $\tau \sim 1$ surface, or the planetary `photosphere', of the dayside atmosphere of WASP-43b. Our observed planet-star flux ratios in the $H$ and $K_s$ bands are shown in Fig.~\ref{fig:sed}, together with the narrow-band photometric data reported by \citet{Gillon2012} at 1.19 $\mu$m and  2.095 $\mu$m. We find that the sum-total of available data can be explained by a blackbody spectrum of the planet with a temperature of ~1850 K, representing the isotherm in the lower atmosphere.  

Our data place good constraints on the day-night energy redistribution in WASP-43b. Assuming a solar abundance composition, we find that models fitting the data require very low day-night energy redistribution ($\lesssim 15\%$), consistent with the finding of \citet{Gillon2012}. Fig.~\ref{fig:sed}a shows a solar abundance model fitting the data with a redistribution fraction of 14\%. However, exploring models with super-solar metallicities allow for slightly higher redistribution fractions, up to $\sim$25 \%. Increasing the metallicity, while keeping the abundance ratios constant at solar values, causes a concomitant enhancement in the molecular abundances, particularly of H$_2$O. The increased molecular abundances cause greater absorption of thermal emission on the dayside atmosphere. The absorbed energy can, in turn, be advected to the night-side atmosphere leading to greater day-night energy redistribution. Fig.~\ref{fig:sed}b shows a model with 5$\times$ solar metallicity and 25\% redistribution which also fits the data. The reason both models, with solar and enhanced metallicities, fit the data is because the data are sensitive only to the lower atmospheric temperature profile which is similar in both the models. On the other hand, the enhanced metallicity model in Fig.~\ref{fig:sed}b clearly shows deeper absorption features compared to the solar abundance model shown in Fig.~\ref{fig:sed}a. Therefore, future observations in the molecular bands should be able to place joint constraints on the metallicity and day-night redistribution of WASP-43b. 

Our data also suggest that a strong thermal inversion in unlikely in the dayside atmosphere of WASP-43b. Even though our data can be fit by a model atmosphere with a strong thermal inversion, such a model would violate energy balance. As discussed above, our data are sensitive only to the temperature structure in the lower atmosphere and insensitive to that in the upper atmosphere where inversions typically form (at pressures below $\sim$0.1\,bar; see Fig~2 of \citealt{MS2009}. With the lower atmospheric isothermal temperature fixed by our current data, the presence of a strong thermal inversion would cause greater emergent energy than that of a blackbody spectrum fitting the data. This is because a temperature inversion causes emission features superimposed on the blackbody continuum of the lower atmosphere as opposed to absorption features formed when the temperature decreases outward monotonically in the absence of a thermal inversion \citep{MS2010}. As such, we find that the presence of a strong thermal inversion would almost violate energy balance by causing greater energy output than the incident energy. However, our data are unable to rule out the presence of a weak temperature inversion, one with only a marginal temperature excess ($\lesssim$100 K) over the isotherm of the lower atmosphere or one where the inversion occurs in regions of low molecular opacity, e.g. for pressures below $\sim$10$^{-2}$ bar.

On the other hand, a thermal inversion in WASP-43b is also less likely on theoretical grounds if molecules such as TiO and VO are the cause of thermal inversions in hot Jupiters~\citep{Hubeny2003, Fortney2008}. We find that the best fitting temperatures of the lower atmosphere as constrained by our data are cooler than the condensation curve of TiO at the same pressures, which would lead to the formation of a TiO cold trap. The TiO/VO condensation together with their gravitational settling, owing to their large molecular mass, would make it unlikely for gaseous TiO/VO to remain aloft in the atmosphere in the significant amounts required to cause thermal inversions~\citep{Spiegel2009}. 

New observations are required to conclusively constrain the presence of a thermal inversion in WASP-43b. Models with and without thermal inversions that fit our data predict distinctly different  planet-star flux ratios at wavelengths with strong absorption features. For example, solar composition models without thermal inversions predict molecular absorption features due to H$_2$O and CO, whereas those with thermal inversions predict molecular emission features \citep{Burrows2008, Fortney2008, MS2010}. Consequently, observations in spectral bandpasses within the molecular bandpasses would be expected to show brightness temperatures lower (higher) than our observed $H$ and $K_s$ brightness temperatures for models without (with) thermal inversions. Such signatures are observable in the {\it Spitzer} IRAC photometric bandpasses at 3.6 $\mu$m and 4.5 $\mu$m, and in the {\it HST} WFC3 bandpass (1.1-1.7 $\mu$m). 

\section{Conclusions} 
\label{sec:conclusions}
We have detected thermal emission from the very compact hot Jupiter WASP-43b in the $H$ and $K_{\rm s}$ photometric bands, using the WIRCam instrument on the CFHT. We determined the planet-to-star flux ratios to be 0.103$_{-0.017}^{+0.017}\%$ and 0.194$_{-0.029}^{+0.029}\%$ in the $H$ and $K_{\rm s}$ bands, respectively. Our measurement in $K_{\rm s}$ band is in agreement with 2.09$\mu$m narrow band measurements by \citet{Gillon2012}. 

Our $H$ and $K_{\rm s}$ eclipse depths are consistent with a single blackbody spectrum with a brightness temperature of $\sim$ 1850~K. This is higher than the expected equilibrium temperature, but in good agreement with predictions from planetary atmospheric models. Due to the low opacity at the wavelengths of our observations we probe the deep layers of the planetary atmosphere where the temperature structure is expected to be isothermal for hot Jupiters. 

Combining our observations with those available in the literature, and comparing with theoretical models, we are able to constrain the day-night energy redistribution in the planet. Our data allow day-night energy redistribution of up to $\sim$25\%, depending on the metallicity in the atmosphere. Our data, combined with considerations of energy balance, also suggest that a strong thermal inversion is unlikely in the dayside atmosphere of WASP-43b, although a weak thermal inversion cannot be strictly ruled out. The current data are insufficient to place stringent constraints on the detailed chemical composition of the atmosphere. 

Follow-up observations using existing ground-based and space-borne facilities will allow for better constraints on various aspects of the dayside atmosphere of WASP-43b, such as its metallicity, presence (or absence) of a thermal inversion, day-night energy redistribution, and molecular composition. In particular, observations with warm {\it Spitzer} at 3.6 $\mu$m and 4.5 $\mu$m will allow to constrain any thermal inversion. On the other hand, observations using the {\it HST} WFC3 instrument would allow to further constrain the water vapor abundance which, in turn, can place stringent constraints on the metallicity and day-night energy redistribution in the planetary atmosphere. Observations of thermal phase curves using {\it Spitzer} and/or {\it HST} will also be able to further constrain the day-night energy redistribution in the planet. Furthermore, ground-based spectroscopic observations and narrow-band photometric observations centered on molecular bands could help constrain the abundances of molecules such as CO and CH$_4$ in the future. 
\section*{Acknowledgments}
This research uses data obtained through the Telescope Access Program (TAP), which is funded by the National Astronomical Observatories, Chinese Academy of Sciences, and the Special Fund for Astronomy from the Ministry of Finance. We thank Karun Thanjavur for the photometric observations at CFHT. WW is supported by the Chinese National Science Foundation grant \#11203035 and \#11233004. NM acknowledges support from the Yale Center for Astronomy and Astrophysics (YCAA) at Yale University, USA, through the YCAA postdoctoral prize fellowship. 

\bibliography{planet_atmosphere}

\label{lastpage}

\end{document}